\documentstyle[epsfig,cite,12pt]{article}

\newlength{\dinwidth}
\newlength{\dinmargin}
\begin{document}

\def\z{${\rm Z}^0$}
\def\ep{e$^+$e$^-$}
\def\ie{{\sl i.e.}}
\def\eg{{\sl e.g.}}
\def\ea{{\sl et al.}}
\def\vrs{{\sl vs.}}

\def\col{Collaboration}
\def\MC{Monte Carlo }
\def\HW{{\sc Herwig} }
\def\OP{OPAL}
\def\DE{DELPHI}

\def\al{\langle}
\def\ar{\rangle}
\def\ka{\kappa}

\def\vs{\vspace*}
\def\hs{\hspace*}
\def\bea{\begin{eqnarray}}
\def\eea{\end{eqnarray}}
\def\be{\begin{equation}}
\def\ee{\end{equation}}
\def\la{\label}
\def\nwp{\newpage}
\def\ct{\cite}
\def\bi{\bibitem}

\def\pT{p_T}
\def\phi{\Phi}
\def\yf{y$$\times$$\phi}
\def\yf{y$$\times$$\phi}
\def\yp{y$$\times$$\pT}
\def\fp{\phi$$\times$$\pT}
\def\3d{y$$\times$$\phi$$\times$$\pT}
\def\lsim{\:{\stackrel{<}{_\sim}}\:}

\date{}

\def\jour#1#2#3#4{{#1} {#2} (19#3) #4}
\def\jourpt#1#2#3{{#1} (19#2) #3}
\def\jourm#1#2#3#4{{#1} {#2} (20#3) #4}
\def\PRp{Phys. Reports}
\def\PRD{Phys. Rev. {D}}
\def\PRC{Phys. Rev. {C}}
\def\PRL{Phys. Rev. Lett. }
\def\AP{Acta Phys. Pol. {B}}
\def\ZP{Z. Phys.  {C}}
\def\EPJ{Eur. Phys. J. {C}}
\def\IJ{Int. J. Mod. Phys. {A}}
\def\NIM{Nucl. Instr. Meth. {A}}
\def\CP{Comp. Phys. Comm.}
\def\TPS{Theor. Phys. Suppl.}
\def\PL{Phys. Lett.  {B}}
\def\NPB{Nucl. Phys.  {B}}
\def\MPL{Mod. Phys. Lett. {A}}
\def\JP{J. Phys. {G}}
\def\NC{Nuovo Cim.}
\def\HEPC{High En. Phys. \& Nucl. Phys.}


\renewcommand{\Huge}{\huge}

\title{\vs{1.5cm}
\bf 
Description of local multiplicity fluctuations\\
and genuine multiparticle correlations\\ 
}
\author{Edward K. G. Sarkisyan{\footnote{ 
        Email address: edward@lep1.tau.ac.il}}\\ 
\small \it School of Physics and Astronomy, 
           The Raymond and Beverly Sackler Faculty of Exact Sciences,\\
\small \it Tel Aviv University, IL-69978 Tel Aviv, Israel
}
\maketitle
\thispagestyle{empty}

\vs{-10.2cm}
\begin{flushright}
{\large TAUP 2619-2000}\\
\date{January 2000}
\end{flushright}
\vs{9cm}


\vs{.7cm}
\noindent
 Various parametrizations of the multiplicity distribution are studied
using the recently published large statistics {\OP} results on
multidimensional local fluctuations and genuine correlations in {\ep}$\to$
{\z} $\to$ hadrons.
 The measured normalized factorial and cumulant moments are compared to
the predictions of the negative binomial distribution, the modified and
generalized versions of it, the log-normal distribution and the model of
the generalized birth process with immigration.
 This is the first study which uses the multiplicity distribution
parametrizations to describe high-order genuine correlations.
 Although the parametrizations fit well the measured fluctuations and
correlations for low orders, they do show certain deviations at high
orders.
 We have shown that it is necessary to incorporate the multiparticle
character of the correlations along with the property of self-similarity
to attain a good description of the measurements.

\nwp
\thispagestyle{empty}


\pagestyle{plain}
\setcounter{page}{1}

\section{Introduction}

The study of the multiplicity distribution of the produced hadrons along
with the analysis of the correlations among them stands in the frontier of
investigations in the area of multiparticle dynamics.
 The multiplicity distribution plays a fundamental role in extracting
first information on the underlying particle production mechanism, while
the correlations give details of the dynamics.
 Whereas the full-multiplicity distribution is a global characteristic and
is influenced by conservation laws, the multiplicity distributions in
restricted phase space domains contributing to the correlations are local
characteristics and have an advantage of being much less affected by
global conservation laws.

In the last decade, study of multiplicity distributions in limited regions
(bins) has attracted high interest in view of search for local dynamical
fluctuations of an underlying self-similar (fractal) structure, the
so-called intermittency phenomenon.
 For review, see Refs. \ct{rev1,rev2}. This phenomenon is seen in various
reactions; however, many questions about intermittency and, in particular,
its origin are still open and further investigations are needed
\ct{rev2,revl}.

Studying local fluctuations, one must remember that the fluctuation of a
given number of particles, $q$, is contributed by genuine lower-order,
$p<q$, correlations.
 To extract signals of these $p$-order correlations, one needs to use
advanced statistical techniques such as normalized factorial cumulant
moments (cumulants)  \ct{mats,cum1,cum2}.
 However, this method requires measurements to be of high statistics (and
of high precision), lack of which leads to a smearing out of high-order
correlations \ct{rev2,cum1}.
 
A real opportunity to extract genuine multiparticle correlations came with
vast amount of multihadronic events collected now at LEP1.
 The statistics available allows one to decrease significantly the
measurement errors and to reveal relatively small effects.
 The first reports representing the studies of genuine correlations in
{\ep} annihilation have just appeared \ct{corD,corO}\footnote
 {Earlier, using correlation (strip) integrals to reduce statistical
errors, multiparticle genuine correlations have been searched for in
hadron-hadron \ct{ci22,cikt} and lepton-hadron \ct{cilh} interactions.
  It is worth to note that except for problems arising due to various
possibilities in defining of a proper topology of particles and a distance
between them (see {\eg} \ct{rev2}), a rather important difficulty in
interpreting results could come from a translation invariance breaking of
the many-particle distributions \ct{ciyl}.
  This will lead to different results on moments/cumulants according to a
variable used because different variables are sensitive to different
hadroproduction mechanisms \ct{ciyl,dBE}.
  For example, in the high-order genuine correlations obtained in Ref.
\ct{ci22,cilh} the study is performed in four-momentum difference squared,
$Q^2$, dependent on Bose-Einstein correlations, whereas in the
pseudorapidity analysis \ct{cikt} no correlations higher than two-particle
ones were found, since (pseudo)rapidity seems \ct{dBE} to be more
``natural'' variable to search for jet formation than {\eg} for
Bose-Einstein correlations.
 }.

{\DE} \ct {corD} has analysed correlations in one- and two-dimensional
angular bins of jet cones, while {\OP} \ct{corO} has performed its study
in three dimensional phase space using conventional kinematic variables
such as rapidity, transverse momentum and azimuthal angle.
 {\DE} has shown an existence of the correlations up to third order, and
{\OP} with increased statistics has, for the first time in {\ep}
annihilation, calculated multidimensional cumulants and has established
sensitive genuine correlations even at fifth order.
 Note that multidimensional analysis carried out in hadronic interactions
shows that the cumulants are consistent with zero at $q>3$ there, while in
heavy-ion collisions non-zero cumulants have been observed at second order
only \ct{rev1,rev2}.

 The genuine correlations measured in {\ep} annihilations are found to
exceed considerably \ct{corD} the QCD analytic description and to indicate
\ct{corO} significant deviations from Monte Carlo (MC) predictions.
 A smallness of genuine correlations predicted by perturbative QCD is seen
also in cumulant-to-factorial moment studies, even when higher orders of
the analytical approximation are used \ct{dc2f,revo}.
 These findings along with the deviations obtained by L3 in its recent
detailed MC analysis of local angular fluctuations \ct{l3af} show particle
bunching in small bins to be a sensitive tool to find differeneces between
parton distributions treated by QCD {\vrs} hadron ones detected in
experiments.
 The study of bunchings seems to be critical to a choice of a more
convenient basis for a suitable approach of multiple hadroproduction.
  
In this paper we compare the intermittency and correlation results from
{\OP} with predictions of various parametrizations, or regularities.
 We consider the negative binomial distribution, its modified and
generalized versions, the log-normal distribution, and the pure-birth
stochastic production mechanism.
 For the first time we examine these parametrizations with high-order
genuine correlations.
 The incorporation of the multiplicity distribution in the study of
correlations provides more advanced information by using various
approximations and models.
 In particular, this study gives more understanding about the structure of
multiparticle correlations, {\eg} their relation to two-particle
correlations.  

All the above listed parametrizations are well-known in multiparticle
high-energy physics for a long time and are used to describe the shape of
the multiplicity distribution, either in full phase space or in its bins.
 Essentially all these parametrizations are sort of branching models and
show an intermittent behavior.
 This feature becomes particularly important in hadroproduction studies in
{\ep} annihilations where parton showers play a significant role and the
hypothesis of local parton-hadron duality is applied to the hadronization
process \ct{revl,revo}.

\section{Normalized factorial moments and cumulants\\
 from various parametrizations}
\la{laws}

There is a variety of models which describe particle production as a
branching process, see {\eg} \ct{revmp}.
 The main prediction of these models is a suitable parametrization for the
multiplicity distribution. 
 Further, more details of the underlying dynamics come from investigations
of the factorial moments and cumulants of the multiplicity distribution
given \ct{rev2}.

One of the most popular parametrization used to describe the data for full
and limited phase-space (basically, rapidity) bins is the {\bf negative
binomial (NB)} distribution, see Refs. \ct{revmp,nbdh,nbdth} for review on
the subject and its historical development.

The NB regularity depends on two parameters: $\al n \ar$ is the average
multiplicity and $1/k$ is the so-called aggregation coefficient which
influences the shape of the distribution.
 For the NB distribution, the normalized factorial moments and cumulants
are governed by the parameter $1/k$ and can be derived \ct{cum2,nbdeq}
from the following formulae,

\be
F_q=F_{q-1}\left(1+\frac{q-1}{k}\right)\,
\la{nbdf}
\ee
and

\be
K_q=(q-1)!\,k\,^{1-q}\, , 
\la{nbdk}
\ee
 respectively.

Fractal properties of the NB distribution have been studied elsewhere
\ct{cum2,enbd,fnbdg,fnbd,fnbds}.
 To demonstrate the fractality of the NB distribution, it has been
proposed to use its generalization \ct{fnbdg} or to study individual
sources obtained by consideration of different topologies, {\eg} number of
jets \ct{fnbds}.
 A good fit by the NB parametrization should yield $k$ independent of
$q$-order, but the limited statistics may also stabilize the
$q$-dependence and hide the (unknown) real distribution \ct{enbd}.
 The bin-size dependence of the $k$ parameter, expected from the bin-size
dependence of the factorial moments and cumulants, Eqs.  (\ref{nbdf}) and
(\ref{nbdk}), is a consequence of the unstable nature of the NB
distribution, {\ie} the convolution of the distributions in two
neighbouring bins does not give the same type of distribution \ct{feller}.
 In the other words, the bin-size dependence of $k$ reflects the fact that
distribution in one bin depends on that in the other bin \ct{nbbd}. 

 In {\ep} annihilations at the {\z} peak \ct{omd,dmdm,dmd,alf,sld} and
lower energies \ct{tasso}, the NB parametrization was found to fail
consistently in describing the multiplicity distribution, either in
rapidity bins or for the full phase space.
 Using {\ep} results on intermittency \ct{DOi}, it was also shown \ct{fmd}
that this parametrization does not reproduce the large fluctuation
patterns, while it is appropriate for phase space bins in which the
fluctuations are sufficiently small.
 Such an effect has been observed in (pseudo)rapidity studies in
other types of collisions \ct{nbdeq,pbi,onbd,my,802} too\footnote
 {Comparing the formation of dense groups of particles in {\ep} and in
hadronic (or nuclear) interactions, one finds noticeable difference of
the fluctuations structure being isotropic (self-similar) in the
former case and anisotropic (self-affine) in the latter one \ct{saf}.
 This is expected to reflect different dynamics of the hadroproduction
process in these two types of collisions.
  }.
 The likely reason is the above noted instability of the NB distribution.
 Indeed, the NB law provides an acceptable description in the central
regions of the multiplicity distribution, away from the tails
\ct{revmp,tasso,alf}.
 In this region, the distribution measured is mostly flat and, therefore,
less sensitive to instability effects.

 Another reason is that the NB regularity underestimates the high
multiplicity tail \ct{fmd,onbd,802,35,n80} which gives the main
contribution to the fluctuations and which is influenced by instabilities.
 For high multiplicities the NB distribution transforms to the stable
$\Gamma$-distribution \ct{feller}.
 This type of distribution was found \ct{omd} to be the most adequate to
describe the multiplicity distribution of the {\OP} data.
 This is in contradiction to the results obtained in nuclear collisions
\ct{802}, where the $\Gamma$-distribution was found to be significantly
inconsistent to reproduce the measurements:
  it underestimates the low-multiplicity parts of the experimental
multiplicity distributions in different rapidity bins, while overestimates
the high-multiplicity tails.
  Again in contrast to {\ep} data, the NB regularity is found to be the
best one to describe small fluctuations in the multiplicity distribution
in nuclear data, and large fluctuations are well reproduced by
two-particle correlations \ct{802,35,n80,wa80,nit}\footnote
 {See preceding footnote.
 }.

Another popular choice for the parametrization of the multiplicity
distribution is the {\bf log-normal (LN) or Gaussian} distribution
\ct{lnd1,lnds}.
 This type of regularity of final state distribution can be obtained by
assuming a scale invariant stochastic branching process to be the basis of
the multiparticle production mechanism.
 The LN distribution is defined by two parameters, the average and
dispersion.
 To describe the data a third parameter has been introduced \ct{lnds} to
take into account an asymmetry in the shape of the full phase space
distribution measured.

In {\ep} annihilation this model has been found to describe successfully
the data for the full rapidity window \ct{omd,alf}. 
 For restricted bins the best agreement between the LN parametrization and
the data has been obtained for very small bins in the central rapidity
region, while for the intermediate size bins the deviation observed is
assumed to arise from perturbative (multi-jet) effects \ct{alf}.
 Compared to the NB, the LN regularity is found to give much better
description which leads to understanding the multiparticle production
process as a scale invariant stochastic branching process.

The fluctuations have been studied in a model of this type in Ref.
\ct{bp}, within the so-called ``$\alpha$-model'' and it was found that in
this particular case the normalized factorial moments obey the recurrence
relation,

\be
\ln F_q=\frac{q(q-1)}{2}\,\ln F_2\,.
\la{lndf}
\ee
 Strictly speaking, this formula connects the standard normalized LN
moments $C_q=\al n^q\ar/\al n\ar^ q$ rather than the factorial moments
$F_q$.  
 The difference between this two types of moments becomes negligible at
large multiplicities which is not the case for small bins. 
 Thus, the results based on the Eq. (\ref{lndf}) with $F_2$ defined by
data can deviate from the true LN regularity predictions, particularly at
small bins \ct{fmd}.

Recently, the {\bf modified negative binomial (MNB)} regularity has been
introduced to correct deviation between the NB parametrization predictions
and the {\ep} and p$\rm {\bar {\rm p}}$ data \ct{mnbd}.
 One finds \ct{mnbdP} for the normalized factorial cumulants of the MNB
distribution,

\be
K_q^-=(q-1)!\,k\,^{1-q}\,\frac{r^q-\Delta^q}{(r-\Delta)^q}\, .
\la{mnbdk}
\ee
 Here, $r=\Delta + \al n \ar / k $ and the superscript minus indicates
that this law is applied for negatively charged particles. 
 The MNB regularity reduces to the NB one if $\Delta =0$, {\sl cf.} Eq.
(\ref{nbdk}).

The MNB parametrization has been found to give an accurate description of
the full phase-space multiplicity distributions in {\ep} annihilation
measured from a few GeV up to LEP2 energies
\ct{dmdm,mnbd,mnbdN,mnbdP2,mnbLEP2,mnbLEP22} and in lepton-nucleon
scattering data in the wide energy range \ct{mnbln}.
 A similar energy dependence of the parameter $k$ has been obtained in
these two types of collisions. 
 Recently, a simple extension of the MNB law has been found to describe
the charged particle multiplicity distributions in symmetric
(pseudo)rapidity bins as well \ct{mnbe}.

In contrast to the NB, the MNB parametrization is shown to reproduce
fairly well the factorial moments and the cumulants of the full like-sign,
{\eg} negatively charged particle phase space from {\ep} data at the
energies ranging from 14 to 91.2 GeV \ct{mnbdfc}.
 The fits of the multiplicity distributions, the moments and cumulants
give rise to $k>0$, $\Delta<0$~\footnote
   { 
 Negative values of the parameter $\Delta$ are interpreted as the
probability $-\Delta$ of intermediate neutral cluster to decay into the
charged or neutral hadron pairs \ct{mnbdN,mnbdP2}.
  Positive $\Delta$ values are also acceptable \ct{mnbd,pbn}, but in this
case $k$ is the maximum number of sources at some initial stage of the
cascading \ct{pbn} changing from one event to another, in contrast with
the fixed $k$ value corresponding to the case of $\Delta<0$ \ct{mnbdN}.
 }
 and $0<r\leq |\Delta |<1$~\footnote
 {
 Nevertheless, at LEP1.5 energy, $\sqrt{s}\simeq 133$ GeV, an inverse
inequality $|\Delta |<r \simeq 0.91$ can be found\ct{mnbLEP2}.
 An increase of the parameter $r$ with increasing energy is allowed but
this complicates the particle-production scenario leading to the MNB
parametrization \ct{chk-cp}.
  },
 that, according to Eq. (\ref{mnbdk}), leads to the cumulants being
negative at even values of order $q$ and positive at odd ones.
 This fact has been utilized to explain the oscillations of the ratio of
the cumulants to factorial moments as a function of $q$ \ct{mnbdP}.

To obtain the quantities for all charged particles one uses the fact that
the number of charged particles produced in {\ep} collisions is twice that
of negative ones.
 Then, the normalized factorial moments and cumulants are given \ct{mnbdP}
by

\be
K_2=K_2^- +{1\over{2\,k\,(r-\Delta)}}\,, \quad
K_3=K_3^- +{3\,K_2^-\over{2\,k\,(r-\Delta)}}\,, \qquad {\sl etc}.
\la{kmnb}
\ee

The stochastic nature of the NB and MNB laws allows to generalize them by
introducing a stochastic equation of the {\bf pure birth (PB)} process
with immigration \ct{pb,pbw}.
 Then, the NB and the MNB distributions can be derived from this birth
process under the appropriate initial conditions, namely, the birth
process with no particles in the initial stage leads to the NB law, while
the MNB distribution is resulted from the birth process with the initial
binomial distribution \ct{pbn}.

For the PB stochastic process one finds 

\bea
\la{pbf}
\nonumber F_q & = & \Gamma(q)\,x^{1-q}\,L^{(1)}_{q-1}(-x)\\
    & = & 1+\frac{q(q-1)}{x}+\frac{q(q-1)^2(q-2)}{2!\,x^2}
           +\frac{q(q-1)^2(q-2)^2(q-3)}{3!\,x^3} 
+ \cdots\\ 
\nonumber     & = &  1+\frac{q(q-1)}{x}
           +q\sum_{i=1}^{q-2}\frac{q-i-1}
           {(i+1)!\,x^{i+1}}\prod_{j=1}^{i}(q-j)^2
\eea
for the normalized factorial moments \ct{pb,pbw,pbi}, and 

\be
K_q=q\,!\;x^{1-q}
\la{pbk}
\ee
 for the normalized cumulants.
 Here, $L^{(1)}_{q}(a)$ is the associated Laguerre polynomial \ct{matf}.

The PB stochastic model has been applied to describe the multiplicity
distribution and its moments in the entire (pseudo)rapidity range \ct{pb}
and in its bins \ct{pbw} in p$\rm {\bar {\rm p}}$ collisions at c.m.
energy ranging from 11.5 to 900 GeV.
 Considering high order moments, it was shown \ct{pbw} that the results of
this approach are close to the NB predictions revealling the stochastic
nature of particle production and, in particular, of the NB model.
 Further analysis \ct{pbi}, which includes intermittency study extending
from {\ep} to nuclear collisions\footnote{
 See footnote 2.
 }, 
 have shown that the data is well reproduced by this model.
 Note that the predictions of this model are systematically below the NB
calculations.

The last form of the multiplicity distribution, considered in this paper,
is the recently introduced {\bf generalized negative binomial}
distribution called the {\bf HNB} distribution due to the type of special
H-function used to derive it \ct{hnb0,hnba}.
 This distribution represents an extension of the NB regularity to the
Poisson-transformed generalized $\Gamma$-distribution by incorporating
some perturbative QCD characteristics. 
 Varying the shape parameter $k>0$, the scale parameter $\lambda >0$ and
the scaling exponent $\mu \neq 0$, one receives special and limiting cases
of the HNB distribution.
 The Poisson and the LN distributions are special cases of the HNB
distribution in the limit of $k \to \infty$ with $\mu = 1$ and $\mu \to
0$, respectively.
 The HNB regularity converges to the NB distribution at $\mu=1$.

 Applied to high-energy data \ct{hnb0,hnba,hnbh,hnbp}, the HNB
distribution has been found to agree with the data, depending on the
parameter $\mu$ values being positive or negative\footnote{
 For $\mu\lsim 0$ the following reparametrization of the HNB density has
been suggested \ct{hnb0,hnba}:
  $(k,\lambda, \mu)\leftrightarrow (p,\sigma, \alpha)$ with
$p=1/\sqrt{k}$, $\sigma=p/\mu$ and $\alpha=\ln \lambda$.
  With these parameters, {\eg} one gets the LN distribution when $p=0$.  
  } 
 ($|\mu|>1$)  or approaching zero, for different types of reactions.
 The HNB regularity with $\mu>1$ and $k=1$ (the Weibull law)  has been
found to describe the data successfully in inelastic pp and p$\rm {\bar
{\rm p}}$ reactions up to ISR energies and in deep-inelastic e$^+$p
scattering at HERA energies in the entire rapidity phase space \ct{hnba}
as well as in its restricted bins \ct{hnbh}.
  The multiplicity distribution from UA5 data of non-diffractive p$\rm
{\bar {\rm p}}$ collisions at $\sqrt{s}=900$ GeV has been shown to be
fitted reasonably well by the LN ($\mu \to 0$, $k \to \infty$) HNB limit
in the full pseudorapidity window and in its symmetric bins.

 In {\ep} annihilations, it was found that HNB describes the data below
the top PETRA energies for $\mu<-1$ and $k=1$ \ct{hnbn}, whereas at high
energies the HNB description favour $\mu\approx 0$ and large $k$
\ct{hnb0,hnba,hnbp}.
 The latter, as well as the above mentioned success of the LN limit HNB
distribution to reproduce the non-diffractive UA5 data, is ascribed to the
LN regularity of the multiplicity distribution\footnote
 {It is worth to mention the remarkable different shapes of the
multiplicity distribution in these two cases, namely, a heavy-tailled form
with a narrow peak in p$\rm {\bar {\rm p}}$ collisions {\vrs} a bell-like
shape in {\ep} annihilation. 
 Nevertheless, recently it was found that the multiplicity distribution in
both types of interactions show the same behaviour expected from the
log-KNO scaling \ct{hlkno}.
 } 
 obtained earlier \ct{omd,alf,lnds} and recently explained by a
renormalization group approach \ct{hnba}.
 For the LEP1 data $\mu $-value lying between $-1.2$ and $-0.6$ and $k
\simeq 20 \div 130$ have been obtained.
 Note that while the error range of the parameter $\mu$ is found to be
small, so that $\mu$ is above $-2.2$ and below zero, the errors for $k$
allow it to vary between $O(10)$ and $O(10^4)$ \ct{hnb0}.
 The situation does not change when the energy increases up to higher than
the {\z} peak or when the multiplicity distribution in rapidity bins
instead that in full phase phase space is considered \ct{hnba}.
 It is interesting that for central rapidity bins, $\mu$ is obtained to be
{\it positive}, $0<\mu<1$, increasing with enlarging the bin size, while
$k$ decreases and is of $O(10)$.
  Recent analysis \ct{hnbp} of the full phase space of uds-quark jet of
the {\OP} data \ct{Ouds} has shown the error bars for the $k$-parameter to
lie between 20 and 420, with the central value at 54.
  The parameter $\mu$ has been found to be about $-0.5$. 

One gets \ct{hnb0}  for the HNB-defined normalized factorial moments, 

\be
F_q=\frac{\Gamma (k+q/\mu)}{\Gamma (k)}
    \left[ \frac{\Gamma (k)}{\Gamma (k+1/\mu)}\right] ^q, 
\la{hnbf}
\ee 
 approaching asymptotically for large $q$ to the $\Gamma$-function of the
rescaled rank, $q/\mu$.

\section{Comparison with OPAL measurements and discussion}
\la{data}

In Figs. \ref{1f} and \ref{1c} we show, respectively, the normalized
factorial moments and the normalized factorial cumulants, measured by
{\OP} \ct{corO} in {\ep} annihilation and compared to a few
parametrizations (lines) and to the MC \ct{mc} (shaded areas).
 The moments are represented in one-, two- and three dimensions of the
phase space of rapidity, transverse momentum and azimuthal angle
calculated with respect to the sphericity axis.

For the NB law we used the second-order factorial moments or cumulants to
compute $k$ and then the factorial moments and cumulants of order $q\geq
3$, according to Eqs. (\ref{nbdf}) and (\ref{nbdk}).
 The resulting quantities are shown by the dashed lines.

From Fig. \ref{1f} one can see that in general the NB regularity
underestimates the measured factorial moments.
 The deviation is more pronounced in one and two-dimensions for number of
bins $M>5$ (in one projection) and $q>3$.
 Note that these our conclusions coincide with those from the analogous
investigations \ct{fmd} of earlier LEP1 results on intermittency \ct{DOi}.
 The better agreement between the parametrization and the data we find for
low-order ($q=3$) moments or for those in three dimensions, the cases when
the NB predictions are within the MC results.
 Nevertheless, even in three dimensions, the NB values are below the data
points at large $M$ (small bin size) and high orders.

The situation becomes more clear when one addresses to the cumulants,
namely the genuine correlations contributing to the fluctuations, Fig.
\ref{1c}.
 From one- and two-dimensional cumulants one can see that not only for
high $q$'s, but even in the case of $q=3$ the correlations given by the NB
model are weaker than those from the data.
 The same is observed for $\pT$, $\yp$ and $\fp$ projections (not shown).
 The NB cumulants lie far away from the measured ones in comparison with
the MC predictions being much nearer to the data.
 Moreover, it is seen that the discrepancy between the data and the NB
results does not begin at some intermediate $M$, as in the case of the
fluctuations, but is visible even at smaller $M$-values (larger bin
sizes).
 This observation agrees with the inadequacy of the NB regularity to fit
the full phase space multiplicity distribution, {\ie} at $M=1$.
 The NB parametrization describes the data reasonably well in three
dimensions, although some deviations are seen for $M> 125$.

Using Eq. (\ref{nbdk}) we have estimated the parameter $k$ as a function
of $M$ at different $q$.
 The parameter is found to decrease with increasing $M$ and to depend
weakly on $q$.
 The values of $k$ lie between $\simeq 0.2\div 2.8$ and $\simeq 5\div8$ at
two extreme $M$ values and vary at fixed $M$ with a change in $q$ and with
the dimensionality of subspaces.
 The higher is the dimension of the subspace, the smaller is the lower
bound of the $k$-range.
 These lower bound values  are almost independent of $q$.
 Conversely, the values of upper bound on $k$ show their $q$-dependence.
 They are about 8 at $q=2$ and about 5 at $q=3$ and 4 regardless of the
subspace dimension. 
 According to the expectations \ct{enbd}, the observed values of $k$ and
its $q$-dependence do not seem to be related to the truncation effect but
rather to small cascades.
  On the other hand, taking into account a long enough cascade at the
{\ep} collision energies considered here, this conclusion confirms that NB
encounters difficulties in a reasonable and consistent description of the
measurements.
  It is interesting to note that the values of $k$ obtained are close to
those found \ct{mnbe} in the MNB-type analysis of the multiplicity
distributions in restricted rapidity intervals in {\ep} annihilations at
the {\z} peak \ct{dmd,alf}.
 
Contrary to the NB, the LN regularity overestimates the data regardless of
the dimensionality or the type of the variable used.
 The dotted line in Figs. \ref{1f} shows how the LN predictions compares
the measurements.
 The cumulants are calculated from the factorial moments using their
interrelations \ct{rev2}.
 The smaller the bin size is, the larger the difference is.
 The LN parametrization describes the data quite well for order $q=3$
only.

These findings are in agreement with the earlier studies of LN fits to
factorial moments in {\ep} annihilation \ct{fmd}.
 The LN distribution overestimates high multiplicities\footnote{
 See, however, \ct{enbd}.},
  which leads to an overestimate of the fluctuations
and genuine correlations as shown here.
 In contrast to the studies of the full-multiplicity parametrization
\ct{omd,alf}, the deviations are found for all bin sizes and not only for
intermediate ones.
 This is in contrast with the above mentioned multi-jet (perturbative)
effects~\footnote
 {A better description of the multiplicity distribution in the case of
smaller number of jets in {\ep} annihilation, namely for two-jet events,
compared to the inclusive sample, has been also obtained by {\OP} using
the NB regularity \ct{omd}.
 }
 and indicates significant contribution of the non-perturbative stage
dynamics, {\ie} soft hadronization, to the formation of fluctuations and
correlations.
 This agrees with recent theoretical studies \ct{rev2}.
 Violation of the Gaussian law of Eq. (\ref{lndf}) have already been
observed in nuclear \ct{my,35,wa80} and hadronic \ct{22fc} interactions.

The predictions of another, the MNB regularity, are shown in Figs.
\ref{1f} and \ref{1c} by the solid lines.
 The cumulants are calculated using Eqs. (\ref{mnbdk}) and (\ref{kmnb}),
while the factorial moments are derived from their relationships with
cumulants \ct{rev2}.
 The parameters $r$ and $\Delta$ have been fixed at the values $r=0.91$
and $\Delta=- 0.71$, the best values found to describe at least the third
order cumulants.
 The only parameter depending on the bin size is $k$ extracted from $K_2$.
 The value of the $\Delta$ parameter is near to that found in multiplicity
studies \ct{dmdm,mnbd,mnbdN,mnbdP2,mnbLEP2}, while the parameter $r$ has
the value $r> |\Delta|$, in contrast to the $r\leq |\Delta |$ inequality
obtained from the analysis of the full multiplicity distributions.
 From the Figures, one can conclude that, in general, with the parameters
obtained, the MNB regularity describes the data well, although
underestimates the latter at small bins and the parameters $|\Delta |$ and
$r$ have an inverse hierarchy.

 The obtained exceeding of the value of $r$ over the $\Delta$-value is
expected if one applies the MNB regularity not to {\it negative}-particle
distributions but to {\it all-charged}-particle ones\footnote
 {See, however, footnote 5.}, 
 see {\eg} DELPHI publication \ct{dmdm}.
 In this case, two effect are contributing: the number of sources
increases with increasing the width of the phase space bin, while a
neutral cluster decay gives 0, one or two particles hit in the given bin
\ct{chk-cp}\footnote
 {It is interesting to note that being limited to like-charged particles,
a study of particle bunching is less dependent on correlations induced by
charge conservation (and partly by resonance production), in addition to
the above-mentioned advantage of such a study to be less affected by the
energy-momentum constraints \ct{chk-cp}.
 }.
 The latter effect is taken into account by implementing an additional
parameter \ct{mnbe}. 
 This is not the case when corrected formulae (\ref{kmnb}) are used,
therefore the change of the inequality between $r$ and $|\Delta |$ has a
physical meaning to be investigated.

The resulted factorial moments and cumulants of the multiplicity
regularity given by the PB process, are represented in Figs. \ref{1f} and
\ref{1c} by the dashed-dotted lines.
 The parameter $x$ is extracted from the second order cumulants $K_2$ and
defines all the higher-order moments and cumulants, given by Eqs.
(\ref{pbf}) and (\ref{pbk}), respectively. 
 For both quantities the PB predictions are seen to lie lower than those
from the above considered parametrizations.
 In all the cases the PB calculations underestimate the data. 
 The difference between the PB predictions and the data is in contrast to
earlier lower-energy {\ep} parametrization of the factorial moments in the
rapidity subspace \ct{pbi}, where the PB process has been found to explain
the data while underestimating higher-order moments.

No curves predicted by the HNB regularity are shown in the Figures, since
we would like to estimate the regions of the parameters $\mu$ and $k$ of
this approach.
 Combined analysis of the factorial moments and cumulants, the latter
being calculated as combinations of the factorial moments \ct{rev2}, show
that, assuming $k$ to be positive,  the $\mu$-parameter is obtained to
be either positive or negative.
 The factorial moments and cumulants are found to be sign-changing
functions of $k$ for negative $\mu$, in particular at $-1<\mu<0$, so that
one can find more than one $k$-region which satisfies the
measurements\footnote
  {Our observation contradicts the property of the factorial moments and
cumulants shown \ct{hnb0,hnbn} to oscillate around zero at $|\mu|>1$ and
not {\eg} at $\mu >-1$.
  This disagreement can be assigned to different regions of the parameter
$k$, found to be large ($k\to \infty$)  in the case of the
full-multiplicity distribution studies \ct{hnb0,hnbn} while having finite
values in our investigation ({\sl vide infra}).
 }.
 In this case we take the largest $k$-value, above which no sign-changing
behaviour is seen and the calculation fits most of the data.
  For $\mu>0$ the factorial moments and the cumulants decrease with
increasing $k$.
 
At fixed $k$, the absolute value of the parameter $\mu$ decreases with the
number $M$ of bins.
 The parameter is found to be limited in the interval $0.3\lsim |\mu|\lsim
1.8$ when one considers the quantities under study of order $q=2,3$.
 For $q>3$, one faces problems reaching the highest measured values of the
moments and cumulants and, in the case of the highest $q=5$ cumulants,
fitting their lowest (negative) values.
  This narrows the interval of $\mu$ down to $0.6<|\mu|<1.6$. 
  The values of the shape parameter $k$ depends on $\mu$, but are found 
to be  $\lsim 30$. 
  The smallest $k$ is obtained to be 1.2 for $\mu<0$ and 0.1 for $\mu>0$.
  The larger the value of $|\mu|$ is, the smaller is the interval of
$k$. 
  For example, if at $|\mu|>1.3$ the values of $k$ are only of a few
units, $\simeq 2\div 5$, then for $0.5<|\mu|<1.0$ these values lie between
2 and 20. 
 The parameter $k$ slightly depends on the order $q$, increasing with $q$
for negative $\mu$ and decreasing for positive ones.  
  This parameter depends also on the number of bins, being a decreasing
function of $M$.
  Note that these two properties are similar to those observed here for
the NB regularity.
       
  Comparing the results obtained to the HNB studies \ct{hnb0,hnba,hnbn} of
the LEP data and, particularly, to that of the uds-quark jets \ct{hnbp}
(see also Sect. \ref{laws}), one can see similarities as well as important
differences.
  Similar to those studies we find that the parameter $\mu$ can be
positive as well as negative and does not exceed 2 in its absolute value.
 Moreover, this parameter shows the same behaviour with bin size, and $k$
tends to obtain large values \ct{hnba,hnbp}.
  The main difference from the HNB studies is that the region $\mu \approx
0$ is excluded by our investigation, so that the LN law is not a suitable
one.
  This conclusion agrees with the {\OP} observation \ct{omd} and the above
discrepancy between the data and the LN predictions.
  It is also worth to notice that 
  (i) {\it negative} $\mu$'s can be also used to describe the multiplicity
distributions in restricted bins, in addition to the positive ones found
recently \ct{hnbp}, and
  (ii) the values of the $k$-parameter are less than 30 and do not tend to
infinity.
  All this shows that to describe the hadroproduction process correctly,
one needs a more complicated scenario to be realized than those leading to
the regularities discussed here, even generalized to the HNB case.
  One could not {\eg}, find $\mu$ to be only in the interval $0<\mu<1$ as
it would expected due to our findings for the LN regularity ($\mu \approx
0$) to overestimate the measured fluctuations and correlations while the
NB law ($\mu \to 1$) underestimates them.

\section{Summary and Conclusions}
\la{sum}

To summarize, variuos regularities of the multiplicity distribution of
charged particles are studied in restricted phase-space bins of {\ep}
annihilation into hadrons at the {\z} peak.
 The study is based on recent high-statistics results \ct{corO} on
multidimensional local fluctuations and genuine correlations obtained with
multihadronic {\OP} sample by means of normalized factorial moments and
cumulants.
  Such parametrizations as the log-normal and negative binomial
distributions, modified and generalized versions of the negative binomial
law, and the generalized stochastic birth process with immigration are
considered.
  For the first time these parametrizations, being most common in the
field of multiparticle production, are examined with genuine high-order
correlations.

  All the parametrizations are found to give a reasonably good description
of low-order fluctuations while they do show deviations for the high-order
fluctuations and correlations, especially at small resolutions.
  Some discrepancy can arise from a bin-size dependence of the measured
multiplicity distribution and from its truncation.
 However, in our consideration, the influence of these effects is
minimized since the models are based on the measured second-order
factorial moments and cumulants, which carry most of the information given
by the multiplicity distribution.
 Moreover, the effects mentioned arise mostly according to low statistics
that is not the case for the data considered here, even at higher orders. 
 To note is also that even with low statistics data, a simultaneous
analysis of multiplicity distributions in different bin sizes and the
corresponding factorial moments, carried out in {\ep} annihilation
\ct{fmd} and in nuclear reactions \ct{35}, does not show any sensitive
influence of the finite statistics to the results.

From the study presented, one concludes that genuine high-order ($q>3$)
correlations have to be taken into account when the hadroproduction
process is modelling, in particular by a multiplicative law for particle
distributions. 
 Indeed, since all the parametrizations used are essentially based on the
average multiplicity and the two-particle correlations, the discrepancies
between the predictions and the measurements indicate {\it multi}particle
character of bunching of hadrons.
 This could be considered as a reason why all the regularities give a good
description of fluctuations and correlations at order $q=3$.
 Our conclusion confirms the {\OP} result from Ref. \ct{corO}, on which we
are based here and which shows the important contributions of
many-particle correlations to the dynamical fluctuations by the
decomposition of the factorial moments into lower-order cumulants.
 This our finding is also in agreement with the observation of {\DE}
\ct{corD} shown the measured angular cumulants not to be reproduced by
small genuine correlations given \ct{dc2f,revo} by perturbative QCD.
   
A self-similar nature of multihadron production is another issue of 
the above study.
 All the regularities result from the particle-production process of the
stochastic nature.
 Therefore their capability to show the intermittent behaviour seen in the
data can be attributed to their branching self-similar nature.
 The better agreement between the regularities and the measurements found
in three dimensions, where QCD cascading is expected to be fully
developed, stresses the essential self-similarity of the particle
production mechanism.
 However, the discrepancies obtained show that a suitable hadroproduction
model seems to be more sophisticated than those giving the
parametrizations considered in the present paper.

\section*{Acknowledgements}
 I would like to thank my colleagues from the {\OP} Collaboration and
particularly those from the Tel Aviv team for fruitful discussions and
help.
 Comments on the manuscript and/or relevant communications from 
 G. Alexander,
 G. Bella,
 M. Biyajima,
 S. Chekanov,
 I. Dremin,
 J. Grunhaus,
 S. Hegyi,
 W. Kittel,
 G. Lafferty,
 S. Manjavidze,
 O. Smirnova and 
 O. Tchikilev
are highly acknowledged. 
Thanks go to J. Zhou for providing me with his Ph.D. Thesis and to M.
 Groys for his assistance.




\textheight=24.2cm
\nwp
\begin{figure}
\vs{6cm}
\hs{1.5cm}
\epsfysize=10.cm
\epsffile[20 150 200 500]{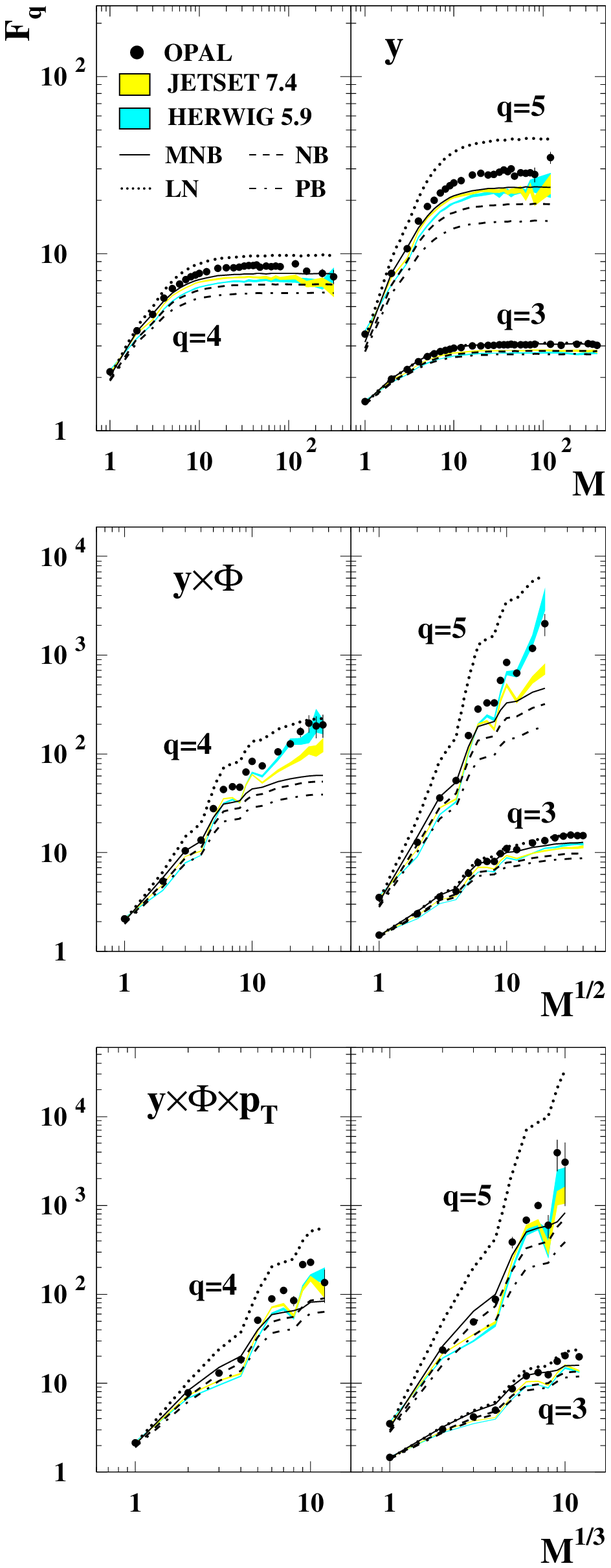}
\vs{2.7cm}
\caption{\it 
Factorial moments of order $q =$ 3 to 5 as a function of $M^{1/D}$, where
$M$ is the number of bins of the $D$-dimensional subspaces of the phase
space of rapidity, azimuthal angle, and transverse momentum, in comparison with 
the predictions of various multiplicity parametrizations and two \MC
models. The data and the \MC predictions are taken from Ref. \ct{corO}. 
}
\la{1f}
\end{figure}

\nwp
\begin{figure}
\vs{6cm}
\hs{1.4cm}
\epsfysize=10.cm
\epsffile[45 150 200 500]{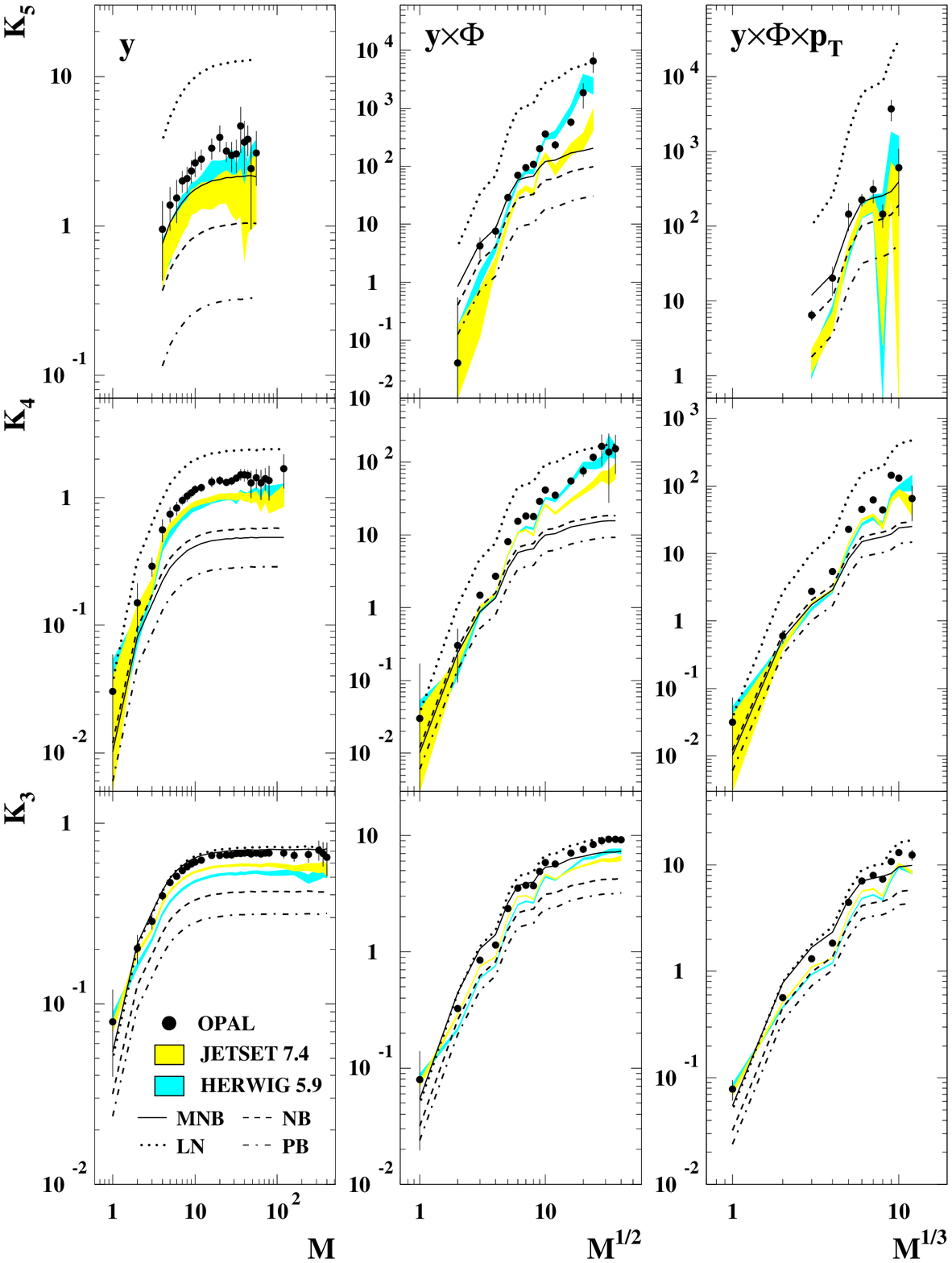}
\vs{2.7cm}
\caption{\it 
Cumulants of order $q =$ 3 to 5 as a function of $M^{1/D}$, where
$M$ is the number of bins of the $D$-dimensional subspaces of the phase
space of rapidity, azimuthal angle, and transverse momentum, in comparison with 
the predictions of various multiplicity parametrizations and two \MC
models. The data and the \MC predictions are taken from Ref. \ct{corO}. 
}
\la{1c}
\end{figure}

\end{document}